\documentclass{egpubl}
\usepackage{eg2019}

 \ConferencePaper % uncomment for Conference submission
 \electronicVersion % can be used both for the printed and electronic version

\ifpdf \usepackage[pdftex]{graphicx} \pdfcompresslevel=9
\else \usepackage[dvips]{graphicx} \fi

\PrintedOrElectronic
\noEGpagenumber

\usepackage{t1enc,dfadobe}
\usepackage{egweblnk}
\usepackage{cite}

\usepackage{amssymb}
\usepackage{amsmath}

\usepackage{graphicx}
\usepackage[usenames,dvipsnames]{color}
\usepackage{subfig}

\newcommand{\figref}[1]{Fig.~\ref{#1}}
\newcommand{\tabref}[1]{Table~\ref{#1}}

\newcommand{\secref}[1]{Section~\ref{#1}}

\definecolor{tbcolor}{RGB}{10,110,210}

\definecolor{dbcolor}{RGB}{210,10,210}

\definecolor{yearacolor}{RGB}{255,0,0}

\definecolor{mbcolor}{RGB}{10,120,10}

\definecolor{hxcolor}{RGB}{255,0,0}

\title[Data-Driven Physical Face Inversion]%
      {Data-Driven Physical Face Inversion}

\author[Kozlov et al.]
{\parbox{\textwidth}{\centering
    Yeara Kozlov$^{1,2}$\hspace{5mm}
    Hongyi Xu$^{3}$\hspace{5mm}
    Moritz B{\"a}cher$^{3}$\hspace{5mm}
    Derek Bradley$^{1}$\hspace{5mm}
    Markus Gross$^{1,2}$\hspace{5mm} 
    Thabo Beeler$^{1}$\hspace{5mm}
    }
    \\
  {\parbox{\textwidth}{\centering 
      $^1$DisneyResearch|Studios\hspace{5mm}         $^2$ETH Zurich \hspace{5mm} $^3$Disney Research
  }
}
}

\teaser{
\begin{centering}
  \includegraphics[width=1.0\textwidth]{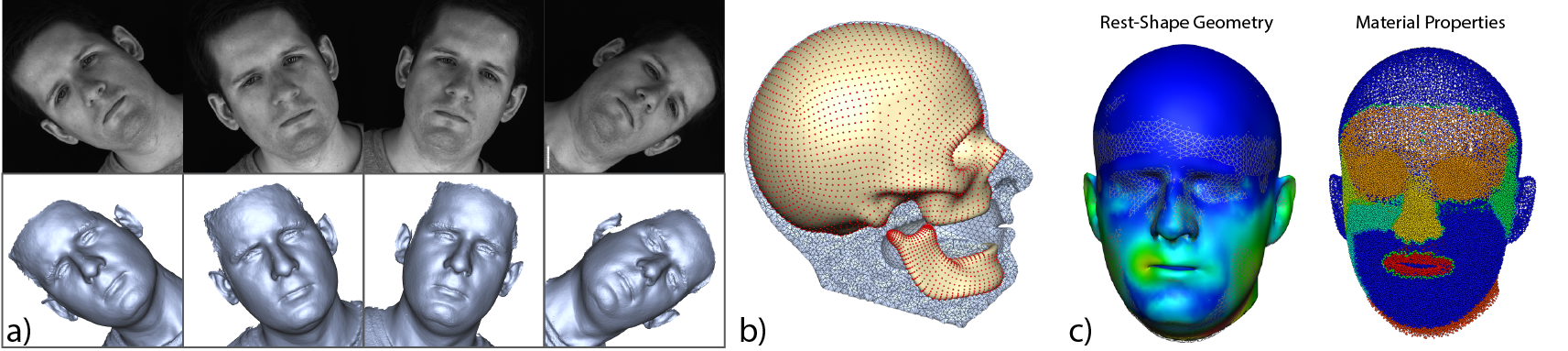}
  \caption{We present a method to capture the physical material properties of real human faces for use in physical simulation for facial animation. From a sparse set of facial surface deformations obtained from a lightweight capture setup (a), and a simple volumetric face model (b), we automatically infer both rest-shape simulation geometry and spatially-varying material properties (red: high Young's modulus; blue: low Young's modulus) for physical simulation (c). \vspace{20pt}} %ugly hack to fix broken spacing. lv
  \label{fig:teaser}
  \end{centering}
}

\begin{document}
\maketitle

\begin{abstract}
Facial animation is one of the most challenging problems in computer graphics, and it is often solved using linear heuristics like blend-shape rigging.  More expressive approaches like physical simulation have emerged, but these methods are very difficult to tune, especially when simulating a real actor's face.  We propose to use a simple finite element simulation approach for face animation, and present a novel method for recovering the required simulation parameters in order to best match a real actor's face motion.  Our method involves reconstructing a very small number of head poses of the actor in 3D, where the head poses span different configurations of force directions due to gravity.  Our algorithm can then automatically recover both the gravity-free rest shape of the face as well as the spatially-varying physical material stiffness such that a forward simulation will match the captured targets as closely as possible.  As a result, our system can produce actor-specific, physical parameters that can be immediately used in recent physical simulation methods for faces.  Furthermore, as the simulation results depend heavily on the chosen spatial layout of material clusters, we analyze and compare different spatial layouts.% both quantitatively and qualitatively \db{remove qualitative if we don't do a dynamic sim comparison}.
\end{abstract}

\section{Introduction}

Creating realistic physical effects for human faces using simulation methods in computer graphics is an extremely challenging problem. On the one hand, linear surface rigs, such as blendshapes, offer intuitive control but are not suited for physical simulation and physical effects have to be added manually. On the other hand, anatomically accurate rigs are hard to create and control, and require high expertise in both human anatomy and CG animation. Recently, simple volumetric rigs have been proposed as middleground \cite{Ichim2016,kozlov2017,ichim2017phace}, which maintain the intuitive control of blendshapes but are also suited for physical simulation. Finding suitable material properties for such rigs is, however, a challenging task, which so far is only possible in a manual trial and error process. And since physical properties change significantly between humans, due to tissue distribution, age and body mass index (BMI), they have to be re-specified anew for every character. In this paper we investigate how to directly estimate the physical properties of human faces from captured data.

One possible approach to this problem would be to measure tissue properties in-vitro for the different tissues that constitute a face, and then distribute these properties assuming some knowledge of the internal tissue distributions. Of course, usually such knowledge is not available, and even in a case where the internal structure of the tissue can be observed, i.e. from MRI or CT data, and when the anatomical structures are known, recovering material parameters for simulation (e.g. stiffness) is still challenging since in-vitro uni-axial strain-stress response measured from thin samples do not directly map to the behaviour of larger tissue structures in-vivo.

We feel a more promising approach is to measure the material properties of human soft tissue in-vivo. While there have been approaches that measure human tissue at a few sparse locations using a force probe \cite{bickel2009capture,pai2018human}, we propose a much more convenient approach that yields dense measurements across the entire face. We capture a subject under a small set of different head poses, which naturally cause the skin tissue to deform due to the known gravity force. From these measurements we can invert both the rest-shape, i.e. the shape of the face in the absence of gravity, as well as material properties, i.e. spatially-varying material stiffness. Combined, these allow to physically simulate the face, yielding results as close as possible to the captured shapes when applying the respective gravitational forces, and extrapolating to physically plausible shapes for other forces. We show that it is essential to represent the face with spatially varying material in order to better approximate the underlying anatomy, and we evaluate several face partition layouts that were presented in the literature. Since the proposed method is agnostic to the layout, we can determine optimal simulation properties for any layout.

To summarize, we present a novel system to estimate rest-shape and spatially varying material properties of real human faces from a few captured 3D scans, making the acquisition and creation of simulation ready facial rigs very practical. While the approach is formulated in the context of human faces, the underlying theory and algorithms are generic and could find application in other areas as well.

\section{Related Work}

Arguably, the most common approach to animate faces is blendshape based facial animation~\cite{lewis2014star}.  However, it is well-known that blendshapes are quite limiting in that they provide only a linear subspace.  Recently, researchers have turned to physical simulation in an attempt to achieve more expressive, non-linear facial animation.

\subsection{Physical Faces}

One of the first methods for physics-based facial animation was presented by Sifakis et al.~\cite{sifakis2005automatic}, who built a detailed face rig consisting of a complete, anatomically accurate muscle structure, created manually from the actor's medical data.  Building the muscle structure for an actor is a time consuming process, therefore Cong et al.~\cite{cong2015} developed an automatic way to transfer a template anatomy to target input faces.  These transferred muscle-based rigs can then be artistically refined by modifying the tracks that muscles follow during activation~\cite{cong2016}.  Ichim et al.~\shortcite{ichim2017phace} also fit a template model of bones, muscles and flesh to facial scans.  Their method succeeds by solving for the muscle activation parameters that best fit the input scans during forward simulation, and thus produces an actor-specific physical face mesh for animation.  A similar approach for full bodies was explored by Kadle{\v c}ek et al.~\shortcite{kadlecek2016}.

Alternative to transferring template anatomy, another approach to avoid the costly setup time of anatomically accurate simulations is to use other representations of the face, for example surface-based~\cite{you2009,ma2012,barrielle2016,barrielle2018} or volumetric rigs~\cite{li2017tvcg,Ichim2016,kozlov2017}.  On the side of surface representations, Ma et al.~\shortcite{ma2012} use a mass-spring system to build a blend shape model that incorporates physical interaction.  You et al.~\shortcite{you2009}, as well as Barrielle et al.~\shortcite{barrielle2016} propose to define forces at the vertices of a face mesh and blend different forces that correspond to different face shapes in order to generate facial animation in a simulation setting, which can even be extended to real-time animation~\cite{barrielle2018}.
While surface based simulation methods are typically easier to setup, they cannot create volume based dynamics such as of soft tissue, which is one of the benefits of volumetric rigs.  On the volumetric side, Li et al~\shortcite{li2017tvcg} propose a method to enrich triangle mesh animations by fitting a tetrahedral mesh, applying physics, and then transferring the secondary motion and collision resolution to the original mesh.  Ichim et al.~\shortcite{Ichim2016}, as well as Kozlov et al.~\shortcite{kozlov2017} propose to more intricately couple blendshape-based facial animation with user-specific volumetric rigs, creating blend-volumes that can be used in a finite-element simulation approach.  The work of Ichim et al.~\shortcite{Ichim2016} allows for several interesting physical effects but does not account for expression-specific dynamics. Kozlov et al. ~\shortcite{kozlov2017} focus on the creation of expression-specific physical effects, but the drawback is that spatially-varying material parameters need to be painted and set manually by an artist for each expression. 
Our method is complementary to Kozlov et al.~\shortcite{kozlov2017} as we aim to {\em automatically} determine the material parameters for a similar volumetric simulation rig.

\subsection{Data-Driven Parameter Estimation}

Our work falls into the category of estimating physical parameters from real world capture data.  Several methods have been presented, outside the focus area of facial animation, for example on the topic of capturing cloth simulation parameters~\cite{bhat2003,wang2011,miguel2012}, optimizing rest shape geometry of elastic objects~\cite{Chen:2014:ANM}, estimating damping properties for dynamics~\cite{Xu2017EDD,DAC2017}, or data-driven modeling of generic soft objects~\cite{wang2015deformation,miguel2016modeling}.

Our work estimates both nonlinear heterogeneous material distribution and rest shape geometry for human faces, under a unified optimization framework of sensitivity analysis. Similar to recent methods~\cite{xu2015interactive,wang2015deformation}, we estimate the heterogeneous distribution of material parameters in a reduced parametric space. However, we parameterize the spatial distribution into several material clusters based on the physical facial structures whereas previous work project the distribution into smooth basis functions without correspondence to real-world data.

Notably, some approaches in the category of data-driven soft object modeling have been designed to measure human tissue.  Bickel et al.~\shortcite{bickel2009capture} fit a stress-strain curve to elastostatic deformation of skin tissue under known forces.  More recently, Pai et al.~\shortcite{pai2018human} measure the mechanical properties of the human body with a new handheld device.  Kim et al.~\shortcite{kim2017} learn an active and passive tissue segmentation and the material parameters for the active tissue for whole-body deformations. Finally, it is worth mentioning that Pons-Moll et al.~\shortcite{pons-moll2015} took a non-physics based approach, but rather a learning approach to model the deformation caused by soft-tissue dynamics in an application of full body animation, by scanning over 40,000 poses of real people.  A similar approach could be used for more expressive facial animation without physical simulation.

\section{Overview}

As illustrated in \figref{fig:teaser}, the input to our approach is a sparse set of aligned facial surface deformations under gravity, 
captured at different orientations of the head with a fixed expression (\secref{sec:data_acq}).
Our approach outputs the simulation mesh geometry at rest (without gravity) and 
spatially-varying material stiffness across the deformable volume of the face,  
such that when we forward simulate the face model under gravity, it will match the target surface deformations. 

We model the deformable face with a tetrahedral mesh and physically simulate the deformation using the Finite Element Method (FEM) with a nonlinear material model (\secref{sec:sim}). 
To obtain the physical parameters for forward simulation, we formulate an inverse optimization to match the captured facial deformations. 
We optimize the unknown physical parameters, by alternating between optimizing the rest shape geometry and material stiffness distribution using sensitivity analysis (\secref{sec:static_inversion}).  
We regularize our inverse problem by aggregating the material distribution into spatial clusters (\secref{sec: matOpt}) and asking for minimal rest-shape geometry changes to our initial guess (\secref{sec:geoOpt}).

In \secref{sect:results}, we analyze the optimized results from different spatial material layouts. 
We validate our optimized physical parameters by simulating to different poses from the capture sequence, which were not part of the optimization targets.

\section{Physically-Based Face Simulation}
\label{sec:sim}

Inspired by \cite{kozlov2017} we do not explicitly model the underlying anatomy but instead treat the entire face as a single volume, with spatially varying material properties that abstract the underlying anatomical complexity. This results in much simpler creation and simulation of a human face, while still achieving physically plausible effects as recently demonstrated in various papers \cite{Ichim2016,kozlov2017,ichim2017phace}.

\begin{figure}
\includegraphics[width=1.0\columnwidth]{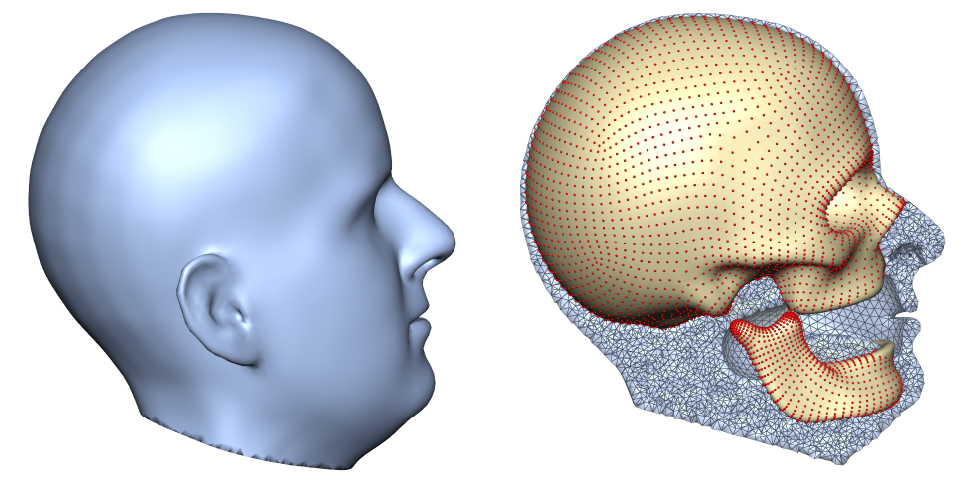}
\caption{The simulation face model consists of a single deformable volume of tetrahedral elements (right, wireframe), conforming to the surface mesh (left), attached to the skull and jaw bone by hard boundary conditions (right, red dots).}
\label{fig:phyiscal_face_model}
\end{figure}

To simulate the deforming facial tissue, we use the Finite Element Method (FEM) simulation of nonlinear elastic materials. The deformable volume is modeled as a tetrahedral mesh, subject to fixed boundary conditions at the skull and jaw (\figref{fig:phyiscal_face_model}).

%\subsection{Material Clusters}
%\yk{Here we describe how we group together tets into clusters}
%See 

%\subsection{Physically Based Simulation}
The simulation model is governed by the underlying hyperelastic material law where the elastic potential energy $W$ is a summation of elemental contribution 
$W_e(\mathbf{X, x, P})$ as 
\begin{equation}
W(\mathbf{X, x, P}) = \sum_e W_e(\mathbf{X, x, P}),
\end{equation}
where $\mathbf{P}$ are the per-element material parameters and $\mathbf{X, x}\in {\mathcal{R}}^{3n}$ are the undeformed and deformed nodal positions, respectively, with $n$ denoting the number of nodes. 
We use a Neo-Hookean material~\cite{bonet2008NCM} to model the nonlinearity in facial deformation, in which case $\mathbf{P}\in{\mathcal{R}}^{2m}$ is a collection of per-element Young's modulus $E_e$ and Poisson's ratio $\nu_e$, with $m$ denoting the number of tetrahedral elements. To improve the simulation robustness to element inversions, we employ invertible FEM simulation using thresholded deformation gradient~\cite{Irving:2004:IFE}.

For quasi-static deformation, the deformed state $\mathbf{x}$ can be determined by balancing the elastic force and external forces as 
\begin{equation}
\label{eq:staticequil}
\nabla_\mathbf{x} W(\mathbf{X, x, P}) = \mathbf{f}_{\text{ext}},
\end{equation}  
where $\mathbf{f}_{\text{ext}}\in{\mathcal{R}}^{3n}$ can be gravity force, contact or any other external loads. 
In section~\ref{sec:static_inversion}, we will use the quasi-static simulation for our inverse problem, where the face is statically deformed under gravity. 

%Following Newton's second law of motion, we can also simulate dynamics by integrating 
%\begin{equation}
%\mathbf{M\ddot{x}} + \nabla_\mathbf{x} W(\mathbf{X, x, P}) = \mathbf{f}_{\text{ext}},
%\end{equation}
%in which $\mathbf{M} \in {\mathcal{R}}^{3n\times 3n}$ is the mass matrix. \tb{assuming this last equation is important later on? Maybe forward reference.}

\section{Static Inversion}
\label{sec:static_inversion}

%\emph{In all derivations, we use the numerator-layout (e.g., an objective gradient is a row vector). We further use total derivatives for implicit dependencies (see supplemental material).}

Given the face model described in \secref{sec:sim}, under gravity the facial tissue would be statically deformed. In this section, we describe how we invert this effect by estimating simulation-ready rest-shape geometry $\mathbf{X}$ and per-element material parameters $\mathbf{P}$ based on observations of the static facial deformation under known gravity forces.  In the following section, we outline a capture procedure and processing pipeline to generate such observations (\secref{sec:data_acq}).

\subsection{Inverse Problem}

Given a deformed configuration and the initial simulation mesh topology, 
we aim at solving for the undeformed vertices $\mathbf{X}$ and material parameters $\mathbf{P}$. 
We formulate our optimization objective as
\begin{gather} 
g(\mathbf{X, P}) = \frac{1}{2} \left\| \mathbf{Sx} (\mathbf{X, P}) - \bar{\mathbf{x}}_s \right\|^2 + \alpha R(\mathbf{X}),\\
\textrm{subject to} \quad \mathbf{P}_l \leq \mathbf{P} \leq \mathbf{P}_u,
\end{gather}
where $\bar{\mathbf{x}}_s \in \mathcal{R}^{3s}$ denote the subset of deformed and observed vertices on the surface and $\mathbf{S} \in \mathcal{R}^{3s\times 3n}$ is the selection matrix to the corresponding observed vertices. $R(\mathbf{X})$ is a regularization term we will use in our rest shape geometry optimization (section~\ref{sec:geoOpt}). 
$\mathbf{P}_l$ and $\mathbf{P}_u$ are the lower and upper bounds for our material parameters respectively.  
Note that the relationship $\mathbf{x}(\mathbf{X, P})$ is non-trivial. Given a new rest configuration or an updated set of material parameters, we first solve a quasi-static forward simulation (Equation~\ref{eq:staticequil}) to get the deformed configuration $\mathbf{x}$ at static equilibrium before we compare the result to the observed vertices $\bar{\mathbf{x}}_s$.

% regularize with a uniform Laplacian formed on a tetrahedral element mesh with known vertex valances but unkown vertex positions.
To minimize this objective with a quasi-Newton method with BFGS, an analytical gradient is desirable,
\begin{equation} 
\left[ 
\begin{array}{c}
\frac{\text{d} g(\mathbf{X}, \mathbf{P})}{\text{d} \mathbf{X}} \\
\frac{\text{d} g(\mathbf{X}, \mathbf{P})}{\text{d} \mathbf{P}}
\end{array}
\right]
= 
\left[ 
\begin{array}{c}
\alpha\frac{ d R(\mathbf{X})}{d \mathbf{X}} + \left(\mathbf{Sx} (\mathbf{X, P}) - \bar{\mathbf{x}}_s\right)^T \mathbf{S} \frac{\text{d} \, \mathbf{x}(\mathbf{X}, \mathbf{P})}{\text{d} \,  \mathbf{X}} \\
 \left(\mathbf{Sx} (\mathbf{X, P}) - \bar{\mathbf{x}}_s \right)^T \mathbf{S}\frac{\text{d} \, \mathbf{x}(\mathbf{X}, \mathbf{P})}{\text{d} \,  \mathbf{P}}
\end{array}
\right].
\end{equation}

However, there is no direct analytical expression for $\frac{\text{d} \, \mathbf{x}(\mathbf{X}, \mathbf{P})}{\text{d} \, \mathbf{X}}$ or $\frac{\text{d} \, \mathbf{x}(\mathbf{X}, \mathbf{P})}{\text{d} \, \mathbf{P}}$ due to the aforementioned implicit relationship $\mathbf{x}(\mathbf{X}, \mathbf{P})$. 
Therefore we propose to use sensitivity analysis to compute the analytical gradient. 

Here for the simplicity of notation, we assemble all the optimization variables into vector $\mathbf{y}=[\mathbf{X}, \mathbf{P}]$.
Starting from Equation~\ref{eq:staticequil}, we take the derivative with respect to the optimization variables $\mathbf{y}$ on both sides, we then get 
\begin{equation} 
- \frac{\partial \mathbf{f}_{\text{ext}}(\mathbf{y})}{\partial \mathbf{y}} + \frac{\partial^2 W(\mathbf{x, y})}{\partial \mathbf{y} \partial \mathbf{x}} + 
\frac{\partial^2 W(\mathbf{x, y})}{\partial \mathbf{x}^2} \frac{\text{d} \mathbf{x}(\mathbf{y})}{\text{d} \mathbf{y}}  \overset{!}{=} \mathbf{O},
\end{equation}
with $\mathbf{O}$ denoting the zero matrix. We note that in our case, $\mathbf{f}_{\text{ext}}$ is the gravity force only. %, which is equal to $\mathbf{M}(\mathbf{X}) [0, -g, 0]^T.$

%At a static equilibrium, we know that  
%\begin{equation} 
%\mathbf{f}(\mathbf{X}, \mathbf{P}, \mathbf{x}(\mathbf{X}, \mathbf{P})) = - \mathbf{f}^T_{\text{ext}}(\mathbf{X}) + \frac{\partial E_{\text{int}}(\mathbf{X}, \mathbf{P}, %\mathbf{x}(\mathbf{X}, \mathbf{P}))}{\partial \mathbf{x}}  \overset{!}{=} \mathbf{o}.
%\end{equation} 
%Otherwise, there would be a descent direction in which we could make further progress when solving the forward problem.
 
%Taking the derivative with respect to the undeformed configuration $\mathbf{X}$, we get 

We could compute $\frac{\text{d} \mathbf{x}(\mathbf{y})}{\text{d} \mathbf{y}}$ by solving the equation systems
\begin{equation}\label{eq:rest_shape_imp_theorem}
\frac{\partial^2 W(\mathbf{x, y})}{\partial \mathbf{x}^2} \frac{\text{d} \mathbf{x}(\mathbf{y})}{\text{d} \mathbf{y}} = \mathbf{b}(\mathbf{x,y}) 
\end{equation}
with right-hand side
\begin{equation}
\label{eq:rhs}
\mathbf{b}(\mathbf{x,y}) = \frac{\partial \mathbf{f}_{\text{ext}}(\mathbf{y})}{\partial \mathbf{y}} -\frac{\partial^2 W (\mathbf{x, y})}{\partial \mathbf{y} \partial \mathbf{x}}
\end{equation}
where $\frac{\partial^2 W(\mathbf{x,y})}{\partial \mathbf{x}^2}$ is the stiffness matrix we use when solving the forward problem with Newton's method.

\textbf{Adjoint Evaluation.} The dimension of the right hand side (Equation~\ref{eq:rhs}) is $3n\times p$, where $p$ is the number of the optimization variables. Evaluating the system~\ref{eq:rest_shape_imp_theorem} directly requires $p$ linear system solves. Even though we can prefactorize the left hand side matrix, this is still computationally expensive when we have a high-dimensional optimization problem. Here instead we use the adjoint method~\cite{cao2003adjoint} to reduce this into a single system solve. 

We plug the explicit expression
%However, it is more favorable to plug the explicit expression
\begin{equation} 
\frac{\text{d} \, \mathbf{x}(\mathbf{y})}{\text{d} \, \mathbf{y}} = \left( \frac{\partial^2 W(\mathbf{x, y})}{\partial \mathbf{x}^2} \right)^{-1} 
\mathbf{b}(\mathbf{x,y})
\end{equation}
into the expression for the gradient,
\begin{equation} 
\frac{\text{d} g(\mathbf{y})}{\text{d} \mathbf{y}} = \frac{d R(\mathbf{y})}{d \mathbf{y}} + \left( \mathbf{Sx} - \bar{\mathbf{x}}_s \right)^T \mathbf{S} \left( \frac{\partial^2 W(\mathbf{x, y})}{\partial \mathbf{x}^2} \right)^{-1} \mathbf{b}(\mathbf{x,y}).
\end{equation}
%where matrix $\mathbf{S}$ selects the vertices from $\frac{\text{d} \, \mathbf{x}(\mathbf{X})}{\text{d} \, \mathbf{X}}$ that correspond to the observed surface $S$.
By introducing a column vector $\boldsymbol{\lambda}$
\begin{equation} 
\boldsymbol{\lambda}^T = \left( \mathbf{Sx} - \bar{\mathbf{x}}_s \right)^T \mathbf{S} \left( \frac{\partial^2 W(\mathbf{x,y})}{\partial \mathbf{x}^2} \right)^{-1}, 
\end{equation}
we can avoid solving several linear equation systems but solve the \emph{adjoint} system 
\begin{equation} 
 \left( \frac{\partial^2 W(\mathbf{x,y})}{\partial \mathbf{x}^2} \right) \boldsymbol{\lambda} =  \mathbf{S}^T \left( \mathbf{Sx} - \bar{\mathbf{x}}_s \right)  
\end{equation}
instead, resulting in the gradient as
\begin{equation} 
\frac{\text{d} \, g(\mathbf{y})}{\text{d} \, \mathbf{y}} = \alpha \frac{d R(\mathbf{y})}{d \mathbf{y}} + \boldsymbol{\lambda}^T \mathbf{b}(\mathbf{x,y}).
\end{equation}

Instead of optimizing all the variables at the same time, we will use \emph{block coordinate descent} for our inverse problem, alternating between material optimization~\ref{sec: matOpt} and rest shape optimization~\ref{sec:geoOpt}, until convergence. 

\subsection{Cluster-based Material Optimization}
\label{sec: matOpt}
For material optimization, we keep the rest shape geometry $\mathbf{X}$ constant and optimize per-element material parameters $\mathbf{P}$. 
As discussed in section~\ref{sec:sim}, we use Neo-Hookean material for our forward simulation 
where the material stiffness is parameterized with Young's modulus $E$ and volume preservation is controlled with Poisson's ratio $\nu.$
However, we note that for different facial tissues, they mostly differ in the Young's modulus but are close in Poisson's ratio with high incompressibility.   
We therefore only optimize heterogeneous distribution of Young's modulus for our inverse problem but use a constant and homogeneous distribution of Poisson's ratio (we use $\nu = 0.43$).

Optimizing element-wise Young's modulus leads to an inverse problem of $m$-dimension, where $m$ is the number of tetrahedral elements and can be very large as the simulation mesh becomes complex. 
In practice, high-dimensional optimization often suffers from slow converge, poor local minima and parameters overfitting. 
Inspired from reduced material optimization~\cite{xu2015interactive,wang2015deformation} which use basis functions of smooth material distribution, here we project our element-wise material inversion problem into a spatial cluster-based optimization. 

Based on anatomical structures, we first create $c$ spatial clusters ($c \ll m$) of material distribution (Section~\ref{sec:materialClusters}). 
The Young's modulus of each tet is then a linear combination of material cluster values, 
\begin{equation}
E_e = \sum_{i = 1}^{c} w^i_e E_c, 
\end{equation}
where $w^i_e$ is the weight to each cluster for element $e$ and $\sum_{i = 1}^{c} w^i_e = 1.$ 
In our case, we assign each interior element exclusively to a single cluster but for elements at the boundary of clusters, we do linear interpolation between the neighboring clusters. %\hx{need to talk about how these weights are assigned.}
By optimizing the material parameters of $c$ clusters only, we solve a much lower-dimensional problem where the analytical gradient to the $i$-th cluster parameter 
$E^i_c$ is calculated as
\begin{equation}
\frac{d g} {d E^i_c} = \sum_{e = 1}^{m} \frac{d g} {d E_e} w^i_e.
\end{equation}

The initial guess to our material optimization is a homogeneous distribution of Young's modulus where $E_c = 1.0$ MPa for all the clusters. We bound the stiffness  
to be within the physical range of Young's modulus by setting $\mathbf{P}_l = 0.001$ MPa and $\mathbf{P}_u = 1.0$ MPa. For our constrained minimization, we use the standard interior point method and BFGS~\cite{Nocedal2006} as implemented in the KNITRO package.
     
\subsection{Rest Shape Geometry Optimization}
\label{sec:geoOpt}
Although one of our observations will be a deformed surface geometry under gravity at a neutral expression (refer to \secref{sec:data_acq}), the undeformed positions of the simulation mesh $\mathbf{X}$ still remain unknown. To solve for the rest shape geometry, we keep the material parameters unchanged and optimize the rest shape $\mathbf{X}$. Note that we need to find undeformed positions for all the nodes including the interior ones, even though only a subset of the surface mesh nodes $\bar{\mathbf{x}}_s$ are observed by our capture system. 

Similar to element-wise material optimization, the geometry optimization is also a high-dimensional problem with $3n$ number of variables. 
However, here we adopt a different strategy to material optimization by regularizing the objective with 
\begin{equation}
R(\mathbf{X}) = W(\mathbf{X}^0, \mathbf{X, P}^0), 
\end{equation} 
where $\mathbf{X}^0$ is the initial guess to the undeformed geometry and $\mathbf{P}^0$ are the some constant material parameters. 
The regularization term $R(\mathbf{X})$ helps maintain a consistent simulation topology by  
penalizing large deviation of the solution to the initial guess with the elastic energy $W$.  

The initial guess to our rest shape geometry $\mathbf{X}^0$ is obtained by forward quasi-static simulation of the volumetric mesh by applying the opposite gravity force. 
We note that our initial guess is reasonably close to our solution, we therefore use gradient descent method with line search for our optimization. 
With the analytic gradient $\Delta \mathbf{X} = \frac{dg(\mathbf{X})}{d\mathbf{X}}$ evaluated from adjoint evaluation, we do a line search along the gradient direction. To avoid simulation mesh topology change from element inversion, we back track the maximum step length $\beta$ such that no element is inverted~\cite{Smith:2015:BPF} and the first Wolfe condition~\cite{Nocedal2006} is satisfied: 
\begin{equation}
g(X^{(k)}-\beta \Delta \mathbf{X} ) \leq g(X^{(k)}) - \gamma \beta \Delta \mathbf{X}^T \nabla g(X^{(k)}),
\end{equation} 
where $\gamma$ is a control parameter (we use $\gamma = 10^{-4}$).

\subsection{Multiple Poses}
In order to robustly estimate the physical properties, we propose to include multiple poses of the same expression under different head rotations and observe the facial deformations under gravity. 
To compute the rest configuration and material parameters from several observations under varying pose orientations, we extend our objective as
\begin{align} 
g(\mathbf{X}, \mathbf{P}) &= \frac{1}{2} \sum_{o} \| \mathbf{S}^o\mathbf{x}^o(\mathbf{X, P}) - \bar{\mathbf x}^o_{s}  \|^2  
%\nonumber
%&+ \sum_o w_{\text{strain}} E_{\text{strain}}(\mathbf{X}, \mathbf{x}^o(\mathbf{X})) \\ \nonumber
% &+ w_{\text{MRI}} \mathcal{R}_{\text{MRI}}(\mathbf{X}, \mathbf{P}, \mathbf{x}_{\text{MRI}}) \nonumber
\end{align}
%where $\mathbf{x}_{\text{MRI}}$ corresponds to the observation $o$ we registered the MRI scan to.
where each $o$ is an observation of the same expression in a different orientation relative to the gravity. 
We note that here instead of transforming the rest shape $\mathbf{X}$ according to the orientations, we 
align our observed targets to a single pose and rotate the direction of gravity based on the relative orientation, and perform the optimization in the canonical frame.
In practice, we also observe better convergence by starting our optimization from a single pose and using the optimized result as the initial guess to our multiple-pose optimization.

\section{Data Acquisition and Preparation}
\label{sec:data_acq}

We now describe a procedure to capture the observations required for inversion, as described in \secref{sec:static_inversion}, for a real human face.

\subsection{Data Capture}
\begin{figure}[ht]
\includegraphics[width=1.0\columnwidth]{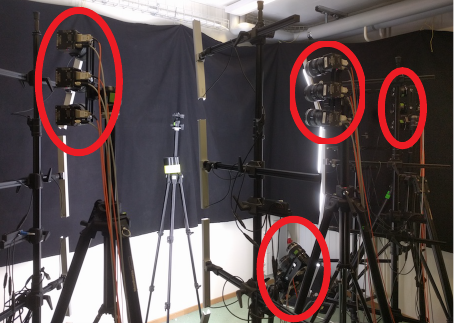}
\caption{Capture setup with four stereo cameras pairs and a tripod standing in for an actor.}
\label{fig:capture_setup}
\end{figure}

Our capture setup consists of four synchronized stereo pairs of machine-vision cameras, capturing at approx. 30fps (see \figref{fig:capture_setup}). 
The method of Beeler et al.~\shortcite{beeler2011} is used to reconstruct geometry from performance sequences of facial motion, and generate a set of meshes in dense vertex correspondence.  The face meshes are stabilized to remove the rigid motion of the head using an approximated skull position, as proposed by Beeler and Bradley~\shortcite{beeler2014rigid}.

\subsection{Actor Performance}
\begin{figure}[ht]
\includegraphics[width=1.0\columnwidth]{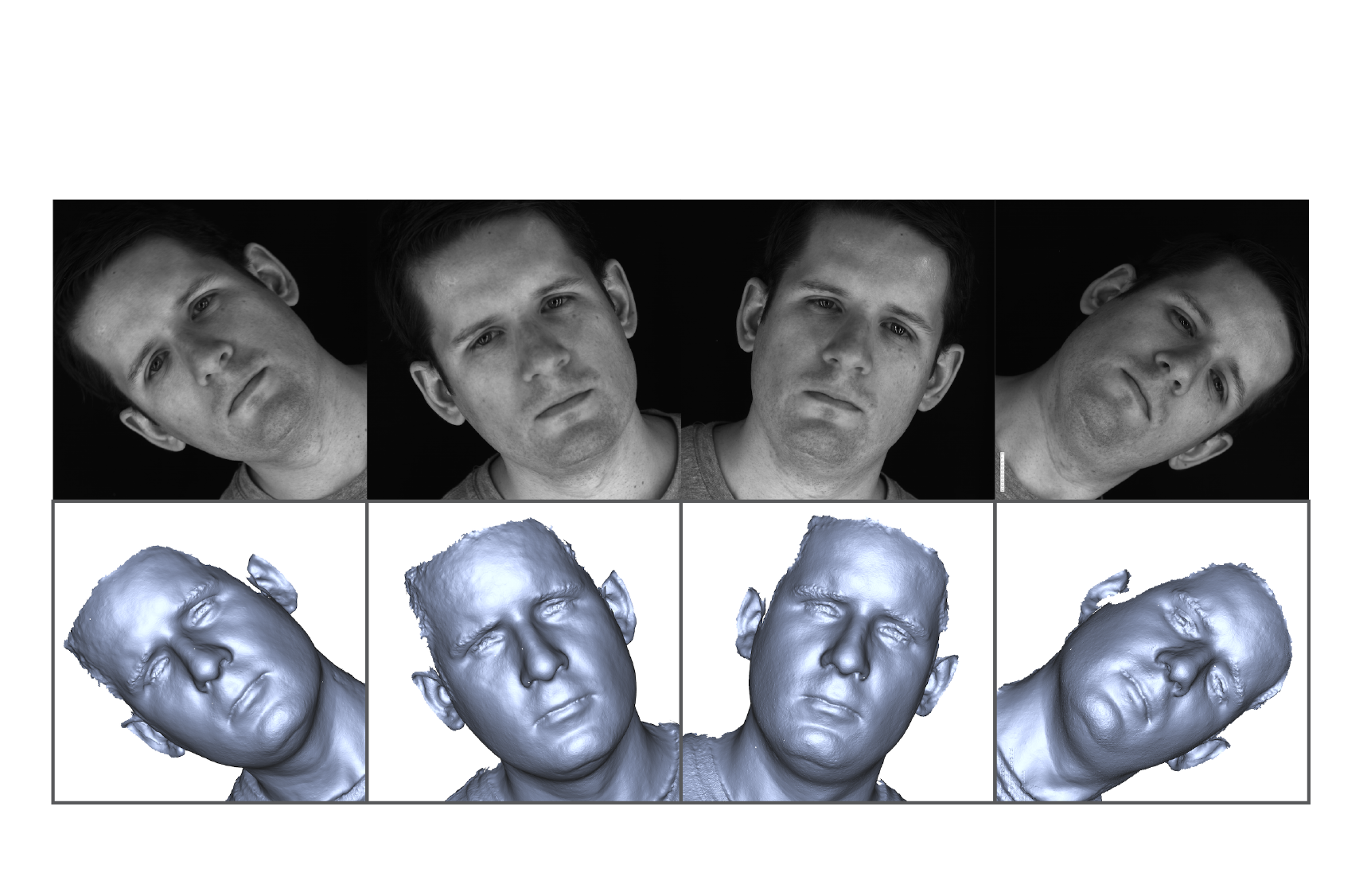}
\caption{Here we show a subset of the captured performance.  The actor tilts his head to each side, providing different tissue deformation due to the known gravitational force.}
\label{fig:capture_data}
\end{figure}

The capture is recorded in a standing position. The actor performs a slow rotation of his head by moving his body and keeping the head position constant relative to their neck. To prevent jaw motion relative to the skull, the actor lightly holds a small teeth guard in their mouth. The slow head rotation prevents any inertia-induced deformation, therefore the observed face deformation is assumed to be quasi-static and only due to the effects of gravity.  A subset of the captured and reconstructed data is shown in \figref{fig:capture_data}.

% The actors held them lightly in their mouth, preventing jaw motion due to gravity change during rotation. We verified this stabilization by looking at the surface alignment of reconstructed, tracked meshes, see section \yk{add section reference and possibly a figure}. \figref{fig:stabilizer}

\subsection{Simulation Model Preparation}
Laplacian deformation is used to deform a closed-surface, generic template face model to one of the target shapes. The underlying skull and jaw bones are estimated using the technique described by Zoss et al.~\shortcite{zoss2018}.

We use Tetgen\cite{tetgen} to generate a surface conforming tetrahedralization of the volume between the face surface and the bone meshes. The correspondences between the resulting volumetric model vertices and the surface meshes of the bones are used to generate the simulation boundary constraints.
The final simulation model consists of 50k vertices and 300k tetrahedral elements. 

\subsection{Target Preparation and Alignment}
\begin{figure}
\includegraphics[width=1.0\columnwidth]{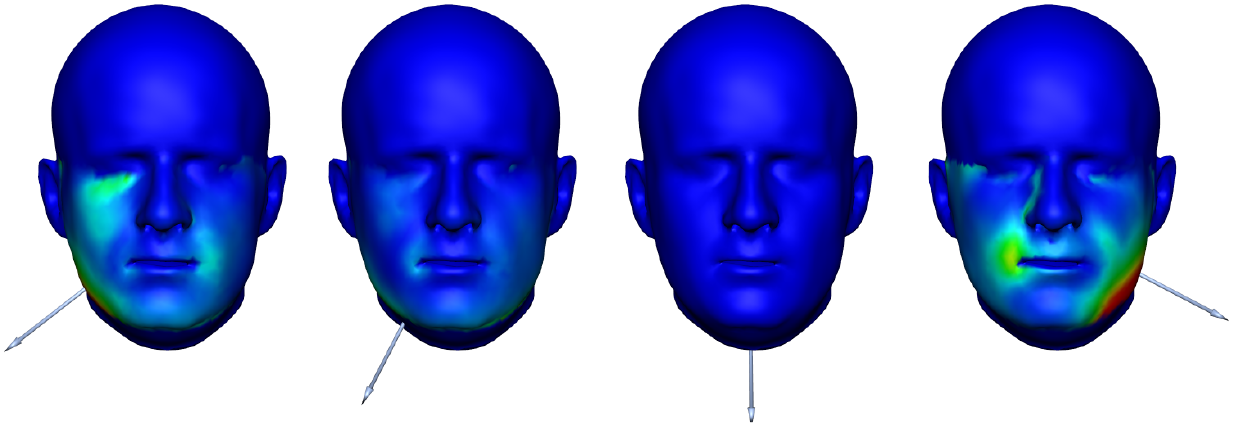}
\caption{Visualization of aligned targets and adjusted gravity directions. The target poses correspond to the neutral, upright pose (third image) and three of the selected poses shown in \figref{fig:capture_data}. The colors encode the vertex displacement from the reference pose, where blue is 0 red is equivalent to 3.5mm.}
\label{fig:data_mod_target_vis}
\end{figure}

From the sequence of tracked meshes, we choose a sparse set of poses to be used as optimization targets (refer to \figref{fig:capture_data}). The selected targets should exhibit a wide span of poses. The effects of the soft tissue deformation exhibited by the face can be easily dominated by any alignment errors, therefore the optimization requires careful treatment of the target poses.  In cases where the automatic stabilization of the meshes is visually insufficient, we manually align the head poses.  The aligned frames are then used to compute a pose-specific adjusted gravity direction, as the optimization is performed in a canonical reference frame. The deformation on the vertices then serves as the observation $\mathbf{\bar{x}}_s$ for our optimization. See \figref{fig:data_mod_target_vis} for visualization of the aligned targets and computed gravity direction. Per target we generate positional constraints for about 1500 vertices of the simulation mesh, selected using a manually specified mask confined to the center of the face.

\subsection{Material Cluster Preparation}
\label{sec:materialClusters}
The segmentation of the mesh into material clusters is done by painting colors on the surface of the volumetric mesh and propagating these colors inwards through the tetrahedral mesh's connectivity. We evaluate two different material cluster layouts; \textit{Layout 1} is based on the material maps from Kozlov et al.\shortcite{kozlov2017}. This layout is comprised of 7 clusters and inherently enforces symmetry since left and right cheeks belong to the same cluster. \textit{Layout 2} is based on the one Tena et al.\shortcite{tena2011interactive} suggested, but we added clusters for the lips as we anticipate those to differ from the surrounding tissue, yielding a total of 19 clusters. The layouts are shown in \figref{fig:res_mat_clusters_stiffness}.

\section{Results}
\label{sect:results}

\begin{figure}
\begin{centering}
\includegraphics[width=1.0\columnwidth]{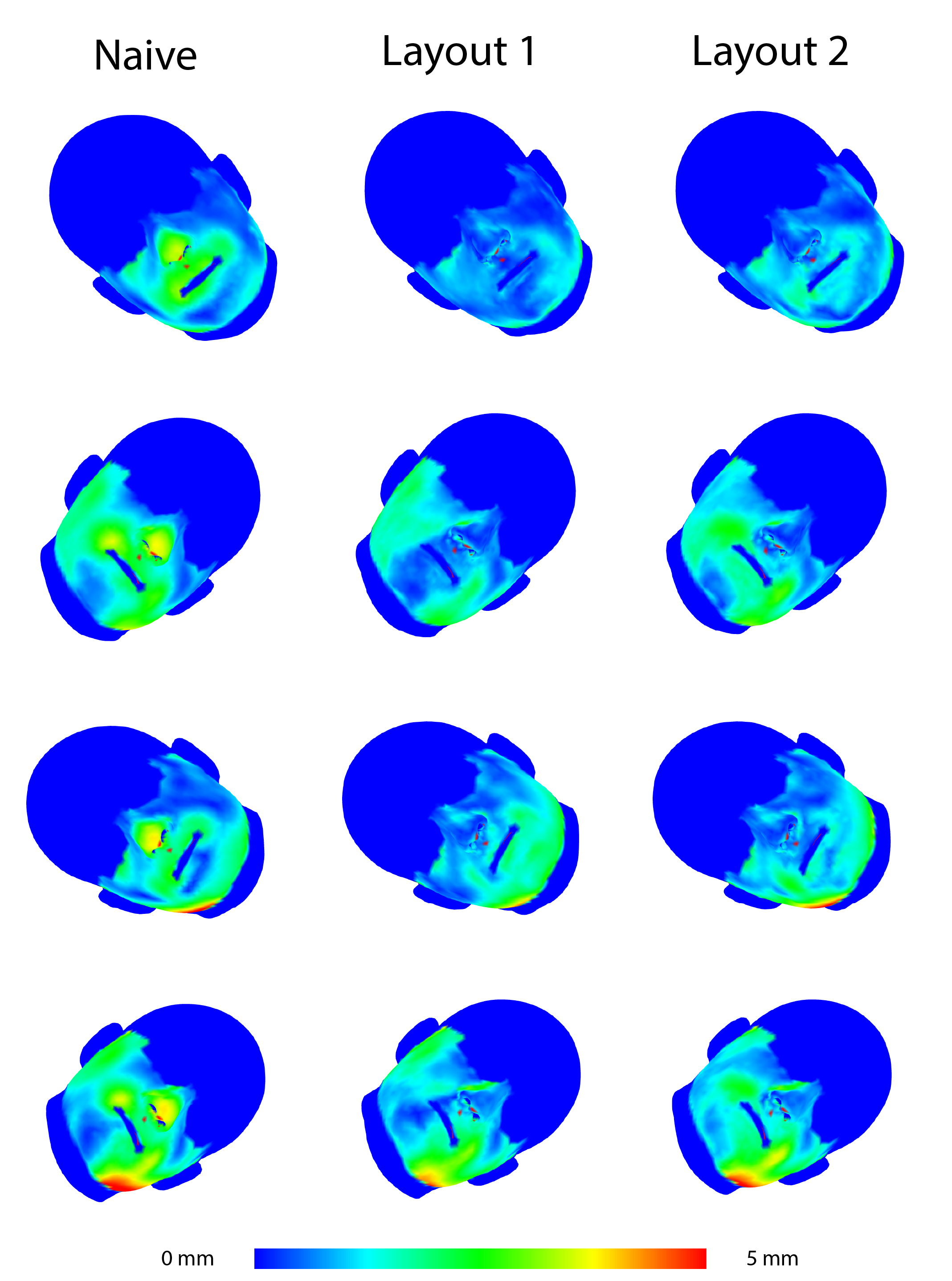}
\caption{Visualization of validation results for different head poses not used during optimization. The naive approach to manually assign a material to the neutral shape yields large errors, in particular for the nose. The chosen material is soft in order to exhibit secondary dynamics during simulation and corresponds to the cheek material for the two other layouts, which have a lower overall error as they can recover spatially varying material properties (\figref{fig:res_mat_clusters_stiffness}) as well as the corresponding res shape geometry (\figref{fig:res_geo_opt_vis}). }
\label{fig:res_validation}
\end{centering}
\end{figure}

%\subsection{Validation}
We validate the results quantitatively by forward simulating to a set of frames where we know the gravity vector and have ground truth surface measurements from the performance capture system. We first simulate using a na{\"i}ve configuration, where a neutral scan is used as rest-shape and we manually determine a uniform material that is soft enough to produce secondary motion and other dynamic effects. While this might seem like a trivial setup, it is probably the most common approach for artists since neither geometry nor material inversion are readily available. Since this treats the entire face as a single material, results are a compromise and, for example, the nose is too soft, leading to large undesired deformations (\figref{fig:res_validation}, left column). Assigning spatially varying materials will remedy this problem, but picking those manually is a challenge in its own. Instead, the proposed system determines the optimal rest-shape in combination with optimal material properties for a chosen layout, which yield a lower error when used in simulation (\figref{fig:res_validation}, center and right columns). The overall error averaged over a total of  136 poses is summarized in \tabref{tab:validation_quant}.

\begin{table}
\small
\centering
\begin{tabular}{|l|c|c|c|}
\hline
Configuration &	Total Error (mm) & Mean Vertex Error (mm) \\
\hline
Naive 		& 	263048	&  1.31 \\ 
Layout 1 				& 	210186 	&  1.05 \\
Layout 2	 	&	212624 	&  1.06 \\
\hline

%Tena	& 93272	 & 0.4669 & 0.0002998715478 \\ 
\end{tabular}
\caption{Validation results. The validation sequence consists of 136 poses, out of which three are used as optimization targets. 
The error is measured as the total absolute point to point vertex error between the simulation result and the tracked mesh.
% Total sequence is the total error for the target vertices over all frames in the sequence. Mean vertex error is the per frame mean vertex error.
}
\label{tab:validation_quant}
\end{table}

\begin{figure}[ht]
\begin{centering}
\includegraphics[width=0.8\columnwidth]{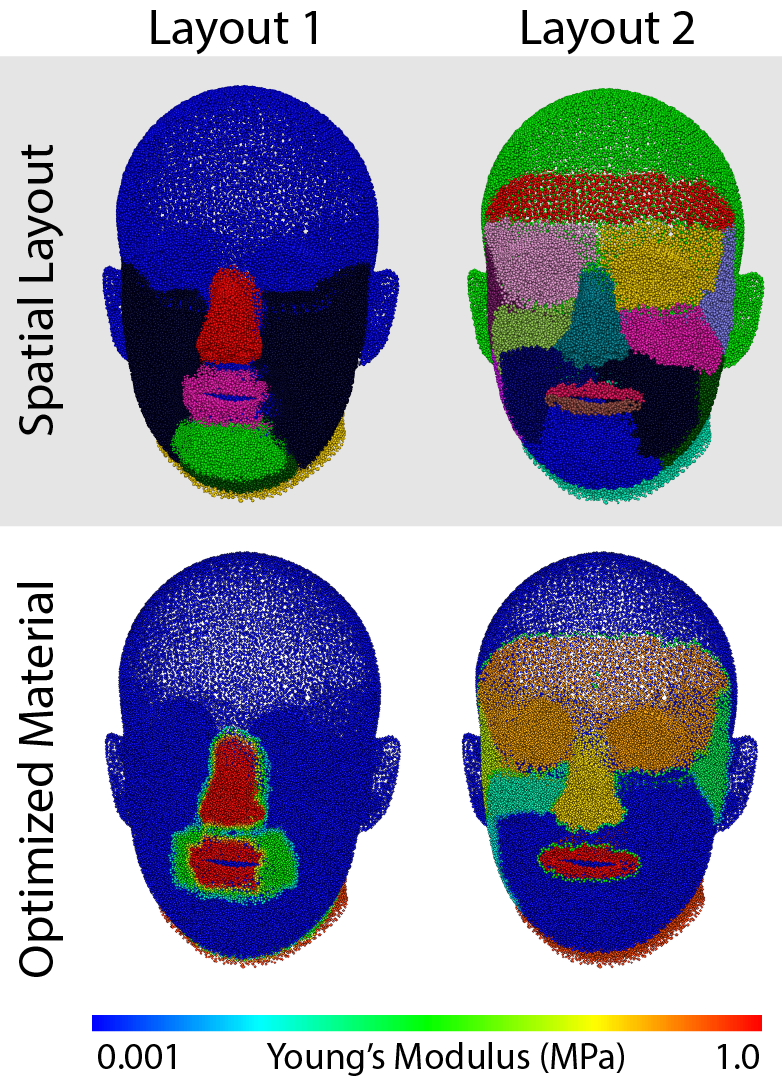}
\caption{Comparison of evaluated layouts and the resulting material stiffness. Please note that even though the layouts differ, the recovered stiffness values for corresponding facial parts are similar.} 
\label{fig:res_mat_clusters_stiffness}
\end{centering}
\end{figure}

\begin{figure}[ht]
\begin{centering}
\includegraphics[width=1.0\columnwidth]{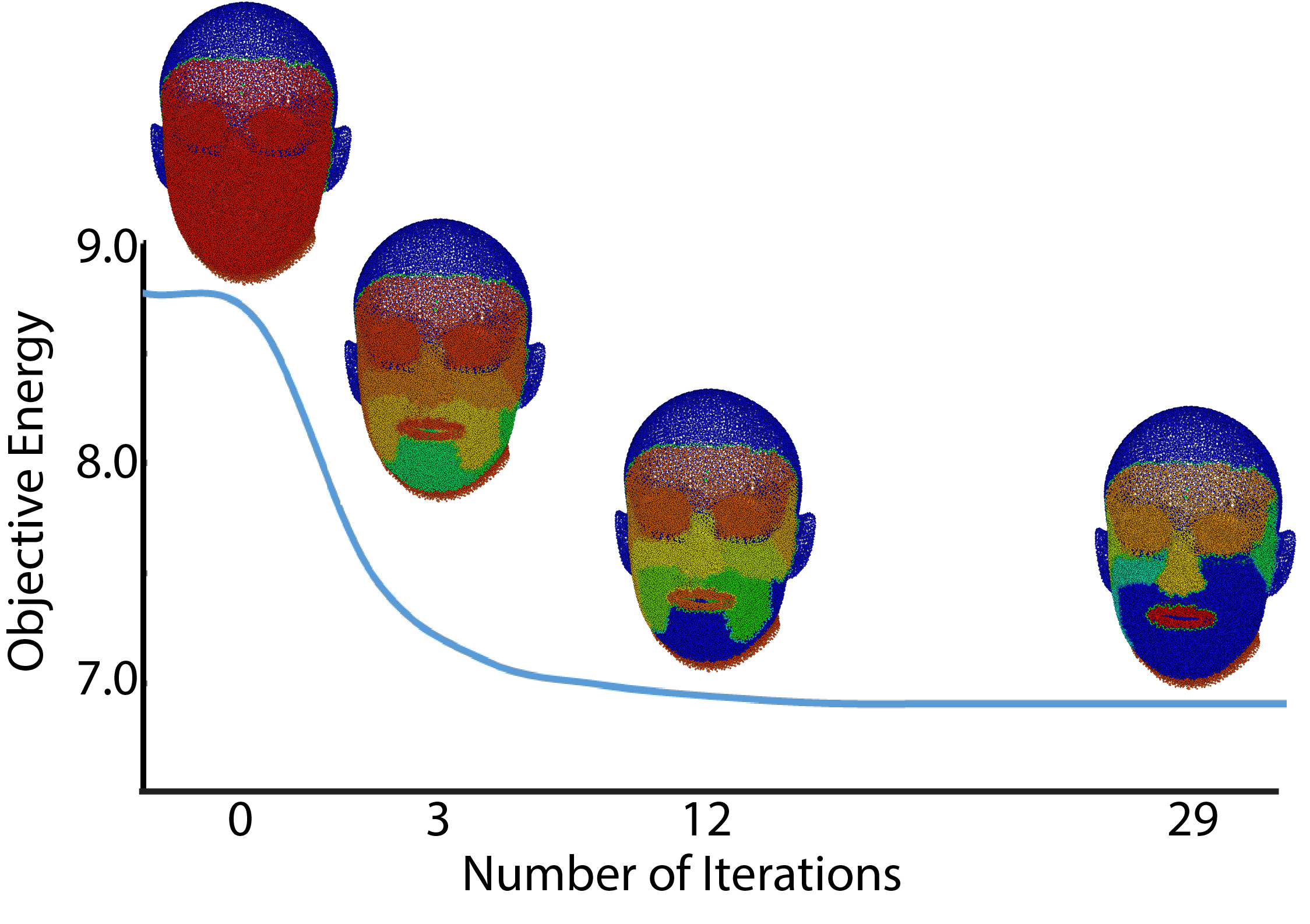}
\caption{Visualization of material optimization convergence for Layout 2 (refer to \figref{fig:res_mat_clusters_stiffness} for the Young's Modulus color scale).}
\label{fig:res_mat_opt_vis}
\end{centering}
\end{figure}

\begin{figure}[ht]
\begin{centering}
\includegraphics[width=1.0\columnwidth]{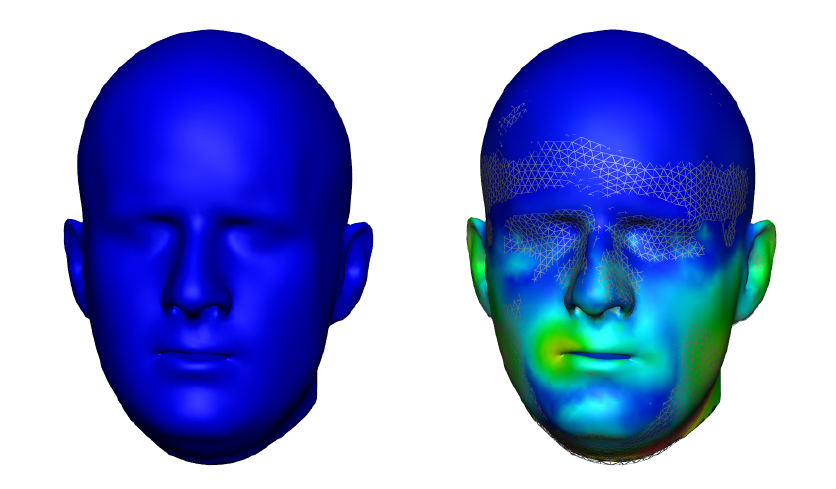}
\caption{Visualization of geometric inversion, which starts from the geometry observed under gravity (left) and optimizes for a different geometry (right) that when used as rest-shape during forward simulation yields again the geometry observed under gravity. The color encodes the distance from original rest shape, where blue encodes 0 and red corresponds to 3mm.}
\label{fig:res_geo_opt_vis}
\end{centering}
\end{figure}

\figref{fig:res_mat_clusters_stiffness} shows the two layouts used in this publication, as well as the resulting Young's Modulus automatically determined by the proposed system. As can be seen, both the layouts yield similar semantic material assignments, i.e. the cheeks come out soft, where nose and lips are stiffer. Starting from an overall stiff material, the optimization successively updates the estimated stiffness per region until convergence as shown in \figref{fig:res_mat_opt_vis}, which is achieved after about 30 iterations. The optimized Young's Modulus stay in the physical range for facial tissues, as reported in literature~\cite{chen1996young}. In an alternating process, we again optimize for the optimal rest shape given the resulting material configuration, which corresponds to the face under zero gravity (\figref{fig:res_geo_opt_vis}). 

\paragraph*{Performance}
The simulation mesh for our face model has $50k$ vertices and $300k$ tetrahedral elements, with $7-19$ material clusters, depending on the layout. 
An unoptimized implementation using five optimization target poses requires up to two minutes for a single evaluation of the function on an Intel i7 3.2Ghz with eight cores, yielding a total runtime of approximately 5 hours. The energies and gradients for the different targets are evaluated in parallel.

\section{Conclusion}

We present a novel method to automatically estimate the physical properties required to simulate real human faces given a small number of aligned facial scans.  As physical simulation for facial animation is gaining popularity, the challenge of tuning the simulation to match a real human face is becoming an increasing roadblock for using simulation approaches in several real-world scenarios (such as VFX and VR). This work represents a large step in the direction of applying simulation methods on real faces. By requiring only a sparse number of input poses, we show that the capture requirements are minimal, and we further study the effects of different spatial layouts for material clustering on the face. 

Our method estimates heterogeneous material properties and rest-shape simulation geometry from multiple poses in a neutral expression. 
In the future, we would like to extend our method to multiple expressions, accounting for material and geometry changes across different expressions~\cite{kozlov2017}. 
Starting from our work, it is also an interesting but challenging direction to obtain physical dynamical properties from sequences of facial performance capture.  
For our rest shape geometry optimization we resize the step length globally to avoid element inversion, which results in robust yet slow convergence. Employing more advanced adaptive approaches~\cite{Zhu:2018:BCQ} could lead to faster convergence and speed up the optimization. 
Furthermore, just like any other data-driven approach, input data noise, 
such as reconstruction error and geometric alignment, has an adverse effect on the optimization quality.
This is aggravated by the fact that the observed deformation due to gravity is rather small, making careful geometric alignment, mouth guard, and proper instructions to maintain the same expression extremely important. In the future we would like to investigate ways to apply stronger forces to observe stronger deformations, while increasing the constraints on the acquisition procedure to reduce input noise further, leading to a better overall signal-to-noise ratio. One avenue along these lines would be to investigate sparse feature-based tracking objectives, compared to our dense point-based optimization targets. 
Lastly, our spatial material layout is just a coarse approximation of the facial anatomical structure and we would like to automatically devise more anatomically accurate material clustering from MRI and CT volume data.

%\subsection{Limitations}
%Our approach comes with a few limitations.
%Some ideas:
%\begin{itemize}
%\item{method is sensitive to geometric alignment and requires that the actor holds the same neutral expression for all target poses.}
%\item the method is also influenced by any deformation or forces created by self collision with the neck.
%\item the geometric optimization is limited in the amount of deformation it can support without inverting tets. we have implicit constraints that are not accounted for directly in the objective or gradient computation. other method may deal with this better.
%\item we do not account for the collision with teeth, lips friction and non rigid skin-bone attachments. this can result in stiffer material parameters in these region. on the other hand, it is also non trivial to model these constraints and set parameters for them.
%\end{itemize}

%\subsection{Future Work}
%Some ideas:
%\begin{itemize}
%\item{extend to multiple different expressions.}
%\item{devise the perfect spatial clustering.}
%\item{capture enough people to build a machine learning approach, or a parametric model from BMI, such that given 1 face scan we can already infer the material properties.}
%\end{itemize}

\bibliographystyle{eg-alpha-doi}
%\nocite{*}
\bibliography{ms}

% \appendix
% \input{appendix}

\end{document}